\documentclass[12pt]{article}
\usepackage{amssymb,amsmath,amsthm,latexsym,amsxtra,graphicx,appendix,epstopdf,feynmf,hyperref,setspace,fix-cm,color,cite,subfig}
\numberwithin{equation}{section}
\onehalfspace
\begin{document} 
\begin{titlepage}
\hbox to \hsize{\hspace*{0 cm}\hbox{\tt }\hss
    \hbox{\small{\tt }}}
\vspace{-1cm}
\hspace{12.4cm}BRX-TH-678

\vspace{2.5cm}
\centerline{\Large{\bf The $a$-theorem for the four-dimensional vector model}}
\vspace{1 cm}
 \centerline{\large 
 {Howard J. Schnitzer\footnote{schnitzr@brandeis.edu}\,}}
\vspace{0.3cm}
\centerline{\it Martin Fisher School of Physics, Brandeis University, Waltham, Massachusetts, 02454, USA}
\vspace{1 cm}
\begin{abstract}
The discussion of renormalization group flows in four-dimensional conformal field theories has recently focused on the $a$-anomaly. It has been shown that there is a monotonic decreasing function  which interpolates between the ultraviolet and infrared fixed points such that $\Delta a=a_{UV}-a_{IR}>0$. In that context Komargodski and Schwimmer showed that $\Delta a$ could be studied by means of dilaton-dilaton scattering. In this paper we examine the $a$-theorem using these methods for a four-dimensional interacting theory: the $O(N)$ vector model, considered to leading order in the $1/N$ expansion and all orders in the coupling constant $\lambda$. 
\end{abstract}
\end{titlepage}
\tableofcontents
\setcounter{footnote}{0}
\newcommand{\be}{\begin{equation}}
\newcommand{\ee}{\end{equation}}
\newcommand{\bea}{\begin{eqnarray}}
\newcommand{\eea}{\end{eqnarray}}
\newcommand{\bb}{\mathbb}
\newcommand{\cl}{\mathcal}
\newcommand{\pr}{\partial}
\def\e{{\rm e}}
\def\cO{{\cal O}} 
\def\tx{\tilde{x}}
\def\half{{\textstyle{1 \over 2}}}
\def\eqn#1{eq.~(\ref{#1})} 
\def\Eqn#1{Equation~(\ref{#1})}
\def\eqns#1#2{eqs.~(\ref{#1}) and~(\ref{#2})}
\def\Eqns#1#2{Eqs.~(\ref{#1}) and~(\ref{#2})}
\section{Introduction}\label{I}
One of the concerns of quantum field theory (QFT) is that of renormalization group flows. In particular for conformal field theories (CFT) if there is a flow of  A $\to$ B, can one also flow from B $\to$ A?. In two dimensions there exists a monotonically decreasing $c$-function, which connects ultraviolet (UV) and infrared (IR) CFT's which is known to be irreversible, such that the number of degrees of freedom decreases as one flows from the UV to the IR CFT's \cite{Zamolodchikov:1986gt}.

Similarly the same question for four-dimensional QFT's is of long-standing interest. A generalization of the two-dimensional discussion uses the fact that the trace of the stress-energy tensor $T_{\mu\nu}$ in a four-dimensional CFT in curved space has $T_\mu^{\;\mu}\ne0$ due to trace anomalies $a$ and $c$, where \cite{Deser:1993yx}
\be
T_\mu^{\;\mu}=aE_4-cW_{\mu\nu\rho\sigma}^2
\ee
with $E_4$ the Euler density and $W_{\mu\nu\rho\sigma}^2$ the square of the Weyl tensor. In this context, Cardy \cite{Cardy:1988cwa} conjectured that 
\be\label{a}
a_{UV}>a_{IR}
\ee
for flows from a UV CFT to an IR CFT. This was originally studied in a number of examples \cite{Osborn:1989td,Jack:1990eb,Schwimmer:2010za}. Holographic models of the $a$-theorem have also been studied in the context of the AdS/CFT correspondence and have provided proof of Cardy's conjecture for these models \cite{Myers:2010xs,Myers:2010tj}.

More recently Komargodski and Schwimmer (KS) \cite{Komargodski:2011vj} proved (\ref{a}) for all unitary RG flows such that there is a monotonic decreasing function which interpolates between $a_{UV}$ and $a_{IR}$. The theorem has been exemplified by several variants of non-interacting theories in various spacetime dimensions \cite{Luty:2012ww,Elvang:2012st,Elvang:2012yc}. There are however few known applications of the theorem to theories with interactions. See also Komargodski \cite{Komargodski:2011xv} and \cite{Schnitzer:2014zxa} for cases with weakly relevant or marginal deformations.

It is the purpose of this paper to apply the KS approach to a particular interacting theory; the O(N) vector model in four dimensions. 

Section \ref{II} provides details of the vector model, computed to all orders in $\lambda$ and lowest order in $1/N$, with emphasis on the pathologies that occur for the model. This involves a calculation of $\Delta a$ using a conformal compensator, which shows that $\Delta a\to+\infty$ for this theory, consistent with a theory without an UV fixed point. A summary of the results appears in section \ref{III}. 

This paper should be considered as a companion of \cite{Schnitzer:2014zxa}. 

\section{$a$-theorem for the vector model}\label{II}
In this section we sketch the argument that the vector $O(N)$ model in four dimensions has $(\Delta a)	\to\infty$. Early references for the $O(N)$ model include \cite{Schnitzer:1974ji,Schnitzer:1974ue,Dolan:1973qd,Root:1974zr,Coleman:1974jh,Kobayashi:1975ev,Schnitzer:1976aq}. A detailed analysis of the model is given in \cite{Abbott:1975bn}, to which the reader is referred for a complete discussion. Here we only select  those features of the model needed to draw the conclusion about the $\Delta a$ for the model.

The renormalized Lagrangian is
\be\label{lag-on}
N^{-1}\mathcal{L}=\frac12\left(\partial_\mu\phi\right)^2+\frac3{2\lambda}\chi^2-\frac12\chi\phi^2-\frac{3\mu^2}{\lambda}\chi,
\ee
where $\phi^a$ belongs to the fundamental representation of $O(N)$ (we use the conventions of (4.1) of \cite{Abbott:1975bn}). The $\beta$-functions for the coupling constant $\lambda$ is
\be\label{beta-on}
(48\pi^2)\beta_\lambda=\frac{\lambda^2}{3}+\mathcal{O}(\lambda^3),
\ee
which is essentially that of \cite{Schnitzer:2014zxa}, with $g^2=0$, with attention to the change of conventions.

The $O(N)$ model is solved to all orders in $\lambda$, and leading order in the $1/N$ expansion \cite{Abbott:1975bn}. The effective potential to this order is
\be\label{effpot}
N^{-1}V(\phi^2,\chi)=-\frac3{2\hat g}\chi^2+\frac12\chi\phi^2+\frac{3\mu^2}{\hat g}\chi+\frac{\chi^2}{128\pi^2}\Big[2\ln\left(\frac{\chi}{M^2}\right)-1\Big],
\ee
where $\hat g$ is the renormalized coupling
\be\label{grenorm}
\frac1{\hat g}=\frac1{\lambda_0}+\frac18\int\frac{d^4k}{(2\pi)^4}\,\frac{1}{k^2(k^2+M^2)},
\ee
and $M^2$ is an arbitrary mass-scale. Analysis of the effective potential shows that it has the following features:
\begin{enumerate}
\item{It has two branches $V_\mathrm{I}$ and $V_\mathrm{II}$, where on branch I the $\chi$-propagator has tachyons, and on branch II, the theory is tachyon free in $1/N$ perturbation theory. The $O(N)$ symmetry is unbroken on branch II.\footnote{Coleman, Jackiw, and Politzer, ref. \cite{Coleman:1974jh} only discuss the unstable branch I.}}
\item{The effective potential has a branch point at a value of $\phi_b^2$ where $V_\mathrm{I}$ and $V_\mathrm{II}$ meet, and $V(\phi^2)$ is complex with no lower bound for $\phi^2>\phi_b^2$. This region is reached from $V_\mathrm{II}$ only by a tunneling transition.}
\item{To leading order in $1/N$ the effective potential satisfies
\be\label{V}
\frac{dV(\phi^2)}{d\phi^2}=\frac12\chi,
\ee
and the renormalized gap equation reads
\be\label{gap}
\chi=\mu^2+\frac16\,\hat g\,\phi^2+\frac{\hat g}{96\pi^2}\,\chi\ln\left(\frac{\chi}{M^2}\right).
\ee}
\end{enumerate}
It will be useful for our discussion of the $a$-theorem for this model to define the gap equation in terms of renormalization group invariant variables \cite{Abbott:1975bn}
\be\label{rho}
\rho(\phi^2)\ln\rho(\phi^2)=-\frac{96\pi^2}{\chi_0}\left(\frac{\mu^2}{\hat g}\right)-\frac{16\pi^2}{\chi_0}\phi^2,
\ee
where
\be\label{chi0}
\chi_0=M^2e^{\frac{96\pi^2}{\hat g}},
\ee
and
\be\label{chi}
\chi(\phi^2)=\rho(\phi^2)\chi_0,
\ee
which shows that $\chi(\phi^2)$ is determined once the two renormalization invariant parameters $\chi_0$ and $(\mu^2/\hat g)$ are specified.

The $O(N)$ model described by (\ref{lag-on}) to (\ref{chi}) violates conformal invariance explicitly. To discuss the implications of the $a$-theorem, we adopt the strategy of \cite{Komargodski:2011vj} by introducing a conformal compensator so as to analyze the RG flow for the model.

In order to consider the vector model in the context of the $a$-theorem, we introduce a compensator field $\Omega$ to restore the conformal invariance explicitly broken by the mass-term in (\ref{lag-on}). The $\lambda\phi^4$ coupling is scale-invariant in the classical theory and becomes irrelevant in the quantum theory. Following \cite{Komargodski:2011vj} we define
\be\label{gamma}
\Omega=fe^{-\tau}=f\left(1-\frac{\varphi} f\right),
\ee
where $\varphi$ is a dilaton field{\footnote{$\phi^a$ is the scalar field and $\varphi$ is the dilaton.}}. The unrenormalized Lagrangian becomes
\be\label{lag-on-phi}
N^{-1}\mathcal{L}=\frac12\left(\partial_\mu\phi\right)^2+\frac12\left(\partial_\mu\varphi\right)^2-\frac12\,\chi\phi^2+\frac32\,\frac1{\lambda_0}\,\bigg[\chi-\frac{\mu_0^2}{f^2}\left(1-\frac\varphi f\right)^2\bigg]^2.
\ee
The renormalized gap equation for the ground state of branch II where $\langle\phi\rangle=0$ \cite{Abbott:1975bn}, i.e. where the $O(N)$ symmetry is unbroken, becomes 
\be\label{rho-phi}
\rho\ln\rho=-\frac{96\pi^2}{\chi_0}\left(\frac{\mu^2}{\hat g}\right)e^{-2\tau}.
\ee
Consider $\delta\mathcal L/\delta\Omega=0$. This has two solutions: $i)$ $\langle\Omega\rangle=f$ which gives $\langle\tau\rangle=0$, and $ii)$ $\langle\Omega\rangle=0$ which yields $\tau\to\infty$. Solution $ii)$ gives a lower effective potential than $i)$. Therefore we analize the consequences of $ii)$ in what follows.
It has been shown by \cite{Komargodski:2011vj} that $\Delta a=a_{UV}-a_{IR}$ can be extracted from dilaton-dilaton forward scattering, i.e. from the term proportional to $s^2/f^4$. Thus we use (\ref{lag-on-phi}) to compute dilaton-dilaton scattering. Since the conformal symmetry is explicitly broken, the dilaton does not appear as internal lines in any diagrams, and one can take $f$ arbitrarily large.

We will not give the details of the calculation of the dilaton scattering amplitude, but only emphasize the necessary features for our discussion. The dilaton fields $\varphi$ only couple to the $\chi$-field, as one observes in (\ref{lag-on-phi}). The $\varphi\,\varphi$ scattering amplitude has contributions from three types of terms: ($i$) bubble diagrams, ($ii$) triangle diagrams, and ($iii$) box diagrams, with $\chi$-propagators the internal lines of the diagrams. The relevant behavior of this classes of diagrams schematically are
\bea
\mathrm{bubbles}:\Big(\frac{s}{f^2}\Big)^2\bigg[\frac{\mu^2/\hat g}{\chi}\bigg]^2,\\
\mathrm{triangles}:\Big(\frac{s}{f^2}\Big)^2\bigg[\frac{\mu^2/\hat g}{\chi}\bigg]^3,\\
\mathrm{boxes}:\Big(\frac{s}{f^2}\Big)^2\bigg[\frac{\mu^2/\hat g}{\chi}\bigg]^4,
\eea
where $\chi=\rho_{\mathrm{II}}(0)\,,\chi_0=m^2$ from (\ref{chi}), with $\chi=m^2$ the physical meson mass (equation (5.1) of \cite{Abbott:1975bn}). 
From equation (5.12) of \cite{Abbott:1975bn} one has
\be\label{br-i}
[1+\ln\rho_{\mathrm{II}}(0)]>2,
\ee
if there is neither a bound-state or resonance, and
\be\label{br-ii}
e>\rho_{\mathrm{II}}(0)>\frac1e,
\ee
if there is a bound-state and resonance. Since the gap equation (\ref{rho-phi}) can be written as
\be\label{gap2}
\ln\rho_{\mathrm{II}}(0)=-96\pi^2\bigg[\frac1{m^2}\left(\frac{\mu^2}{\hat g}\right)\bigg]e^{-2\tau},
\ee
we require
\be\label{limit}
\frac1{m^2}\left(\frac{\mu^2}{\hat g}\right)\to-\infty,
\ee
for the limit $\tau\to\infty$ to be consistent with the gap equation and absence of tachyons. 
Therefore the box diagrams dominate the dilaton-dilaton scattering amplitude.
In slightly more detail, we find the $\varphi\,\varphi\to\varphi\,\varphi$ dilaton amplitude from the box diagrams give
\be\label{delta-a-on}
\Delta a\sim\left(\frac{\mu^2/\hat g}{m^2}\right)^4\left(\frac1{96\pi^2}\right)^4[1+\ln\rho_{\mathrm{II}}]^4.
\ee
Therefore (\ref{limit}) implies that
\be\label{delta-a-infty}
\Delta a=a_{UV}-a_{IR}\to\infty,
\ee
for the vector model computed to all orders in $\lambda$ and leading order in $1/N$. This is consistent with findings \cite{Schnitzer:2014zxa} for the gauged vector model where $\Delta a\to\infty$ if the flow behaves as $y(s)>y_+(s)$, where $y(s)=\lambda(s)/g^2(s)$, which is also true for the $g^2\to0$, $\lambda$ fixed limit, i.e. for the ungauged vector model.
\section{Conclusions}\label{III}
We have studied $\Delta a=a_{UV}-a_{IR}>0$ for the four-dimensional O(N) vector model by means of the dilaton-dilaton scattering, using the method of Komargodski and Schwimmer \cite{Komargodski:2011vj} , with the result that $\Delta a\to\infty$. This is consistent with the analogous result for the gauged vector model when 
$y(s)=\lambda(s)/g^{2}(s)>y_{+}(s)$. See  \cite{Schnitzer:2014zxa} for details. This implies that the four-dimensional vector model has "infinitely many" degrees of freedom in the UV, as there is no ultraviolet fixed point, which suggests the need for an ultraviolet completion of the model {\footnote{We thank Matt Headrick for this remark.}}.

In section 8 of \cite{Abbott:1975bn} it is shown that there is a non-trivial UV zero of the beta-function at
\be\label{lamb-N}
\frac{\lambda^{*}}{48\pi^2}=\frac{N(N+8)}{(N+42)},
\ee
which taken literally renders $\Delta a$ finite. Unfortunately (\ref{lamb-N}) is outside the domain of validity of the 1/N approximation. \textbf{Note added:} Recently Shrock \cite{Shrock:2014zca} has shown that the theory does not exhibit robust evidence of a UV zero up to n=5 loops.
\section*{Acknowledgements}
We wish to thank Steve Naculich for his participation in the early phases of this project. We are appreciative of Matt Headrick and Albion Lawrence  for helpful conversations and comments. HJS is supported in part by the DOE by grant DE-SC0009987. We thank Ida Zadeh for conversations, and Isaac Cohen for his assistance in preparing this manuscript.
\bibliographystyle{utphys}
\bibliography{list-vect}
\end{document}